# DNS of the very large anisotropic scales in a turbulent channel

Juan C. del Alamo[1] and J. Jiménez [1,2]

[1]School of Aeronautics, 28040 Madrid, Spain.
[2]Center for Turbulence Research, Stanford University, CA 94305, USA.

Contact address: juanc@torroja.dmt.upm.es

## 1 Introduction

The large structures in the outer layer of turbulent wall flows are of great physical importance, because they contain a substantial fraction of the streamwise kinetic energy and of the Reynolds stresses [1]. Nevertheless, the organization of the outer region of wall turbulence has historically received less attention than that of the inner region, and as a consequence, it is still the subject of many open experimental and theoretical questions.

In order to address some of those questions, we have performed direct numerical simulations of the turbulent incompressible flow in plane channels at Reynolds numbers $Re_\tau = 185 - 550$ (based on the channel half width h and on the friction velocity). The numerical box is $8\pi h \times 4\pi h$ long in the streamwise (x) and spanwise (z) directions at $Re_\tau = 550$, and even larger ($12\pi h \times 4\pi h$) at $Re_\tau = 185$. Although there are already in the literature computations at comparable Reynolds numbers [2], this is the first simulation in a numerical box large enough not to interfere with the most energetic structures in the flow [3]. The code is fully spectral, using Fourier expansions with dealiasing in the homogeneous directions and Chebyshev polynomials in the wall normal direction. The resolution of the collocation grid is $\Delta x^+ = 8.9$, $\Delta z^+ = 4.5$, $\Delta y^+ < 6.7$.

This paper presents some of the results of this simulation, focusing on the statistical description of the size of the large structures of the streamwise velocity (u) and on its scaling with Reynolds number.

## 2 Results

Figure 1 displays linearly spaced isocontours of the premultiplied two-dimensional energy spectra $\varphi_{uu} = k_x k_z E_{uu}(\lambda_x, \lambda_z, y)$ of the streamwise velocity, as functions

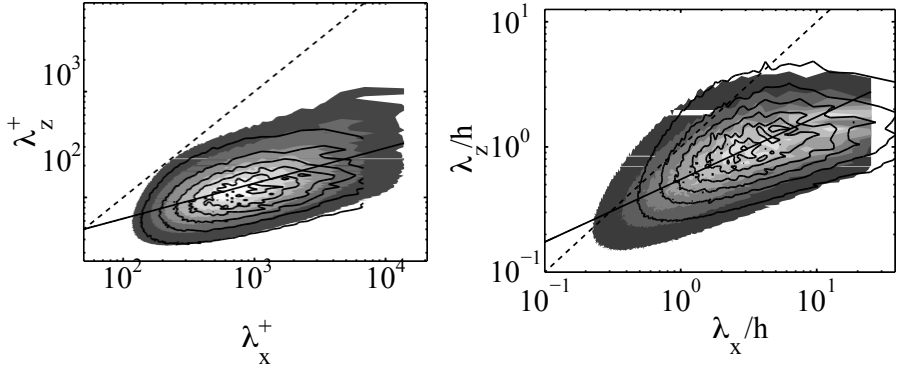

Figure 1: Premultiplied two-dimensional spectra $\varphi_{uu}$ of the streamwise velocity as functions of $(\lambda_x, \lambda_z)$. (a) Wall units, $y^+ = 15$; (b) Outer units, $y = 0.3h$. Shaded contours, $Re_\tau = 550$, line contours, $Re_\tau = 185$. $\cdots\cdots$, locus of 2-D isotropic structures, $\lambda_z = \lambda_x$; ———, (a), $\lambda_x^+ \sim (\lambda_z^+)^3$, (b), $\lambda_x y = \lambda_z^2$.

of the wavelength vector $(\lambda_x, \lambda_z) = (2\pi/k_x, 2\pi/k_z)$. Note that

$$(u'^2) = \int_0^\infty \int_0^\infty \varphi_{uu}(\lambda_x, \lambda_z, y)\, d(\log \lambda_x)\, d(\log \lambda_z), \qquad (1)$$

where $(\cdot)$ denotes averaging along the homogeneous directions and time, so that this figure expresses how much streamwise energy is contained in structures of length $\lambda_x$ and width $\lambda_z$. The shaded contours come from the simulation at $Re_\tau = 550$, while the line contours are from the one at $Re_\tau = 185$. The two wall distances ($y^+ = 15$, Fig. 1(a); $y = 0.3h$, Fig. 1(b)) are representative of the inner and the outer layers of the flow.

In the wall region the spectra of the streamwise velocity peak around $\lambda_x^+ \approx 700$, $\lambda_z^+ \approx 100$, which is the size of the buffer-layer streaks. Inner scaling produces a good collapse of the spectra everywhere in the $(\lambda_x, \lambda_z)$ plane, except in the upper right corner, which corresponds to the location of the large energetic structures of the outer-layer. The u spectra in this region lie approximately along the power law

$$\lambda_x^+ \sim \lambda_z^{+\,3}, \qquad (2)$$

implying that, while the structures of the streamwise velocity become wider as they become longer, they also become more elongated, since they progressively separate from the spectral locus of two-dimensional isotropy.

In the outer region the length and width of the energy-containing eddies of u scale in outer units, as seen in Fig. 1(b). The spectra peak at $\lambda_x = 4h$, $\lambda_z = h$, but note that there is still an important fraction of streamwise energy contained in structures longer than $4h$.

The spectra of v and w, not shown here, are everywhere shorter and more isotropic than those of the streamwise velocity.

An important feature of the u spectra in this region is that they collapse reasonably well around the power law

$$\lambda_x y = \lambda_z^2, \qquad (3)$$

wich is the solid line in Fig. 1(b). A possible explanation for this power law is that the structures of the streamwise velocity are the decaying wakes of approx- imately isotropic smaller v and w structures. They decay under the action of an eddy vicosity $\nu_T$ to diameters of order $\lambda_z$ in times of order $\lambda_z^2/\nu_T$, leaving 'wakes' in the streamwise velocity whose length is

$$\lambda_x \sim U_b \lambda_z^2 / \nu_T, \qquad (4)$$

if we assume that they are convected at a velocity of the order of the bulk velocity. The choice of a constant advection velocity implies that necessarily the large structures feel the wall, or at least the local shear, since velocity itself is not a Galilean invariant. How this behaviour can be reconciled with the different power law observed near the wall (see Eq. 2) is briefly discussed in [4].

Up to now, we have used the spectral representation of the streamwise veloc- ity in order to obtain information about its organization in (x, z) planes, making use of the flow homogeneity in those planes. However, y is not an homogeneous direction, so we need another approach to describe the structure of u across the channel width. For this reason we introduce the two-point autocorrelation function $R_{uu}(r_x, r_z, y, y')$. Besides, in order to discriminate the organization in y of the energetic scales of the outer region, we have restricted (cut-off filtered) $R_{uu}$ to structures longer and wider than $\lambda \approx 0.4h$.

Figure 2 represents the filtered autocorrelation coefficients of u in the (x, y) plane

$$\sigma_{uu}(r_x, r_z, y, y') = \frac{R_{uu}(r_x, r_z, y, y')}{R_{uu}(0, 0, y, y) R_{uu}(0, 0, y', y')} \leq 1, \qquad (5)$$

at $y' = 0.3h$ as functions of y and of the streamwise separation $r_x$. The shaded contours come from the simulation at $Re_\tau = 550$, while the line contours are from the one at $Re_\tau = 185$ Although the no-slip boundary condition imposes that $R_{uu}$ has to vanish at y = 0, this is not the case for $\sigma_{uu}$, which is therefore easier to interpret near the wall than the former. The figure shows that the large structures of u are also tall, reaching both the center of the channel and one of the walls, where $\sigma_{uu}$ takes values of the order of 0.3. This relatively high level of autocorrelation at y = 0 suggests that the incomplete scaling observed Fig. 1 may be due to the penetration of the large outer structures into the wall layer.

## 3 Conclusions

We have performed the first direct numerical simulation of turbulent channel flow using both a computational domain large enough to capture the largest structures in the outer flow and a Reynolds number high enough to observe some separation between those structures and the ones in the near-wall region.

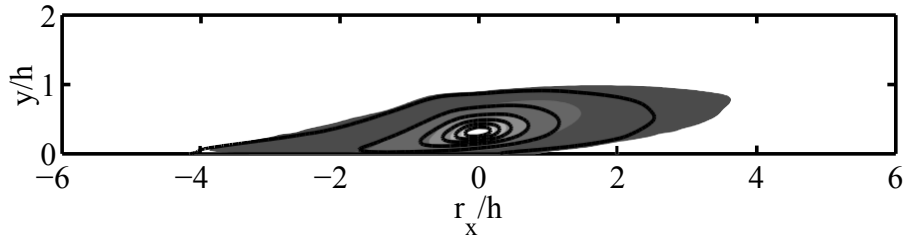

Figure 2: Autocorrelation coefficients $\sigma_{uu}(r_z = 0, y' = 0.3h)$ of the streamwise velocity, as functions of $(r_x, y)$, and restricted to structures with $\lambda_x, \lambda_z > \lambda \approx 0.4h$. Shaded contours, $Re_\tau = 550$, line contours, $Re_\tau = 185$. Outer units.

The results show that there are very large elongated structures in the outer region of turbulent channel flow whose length, width and height scale with h.
We have suggested that they can be understood as the wakes left by compact isotropic structures decaying under the action of an eddy viscosity as they are convected by the mean flow. The results also indicate that these large structures are also very tall, and that they reach the walls, which would help understand the Reynolds number dependence in the scaling of $\langle u'^2 \rangle$ in the near-wall region.

This work has been supported in part by the Spanish CICYT contract BFM2000-1468 and by ONR grant N0014-00-1-0146. The simulation has been run at the CEPBA/IBM center at Barcelona, with time that has been donated by IBM and by the U. Politecnica de Catalunya.